\documentclass[pra,showpacs,twocolumn,amsmath,amssymb,amsfonts,superscriptaddress,letterpaper]{revtex4-1}

\usepackage{graphicx}
\usepackage{dcolumn}
\usepackage{bm}
\usepackage{color}

\begin{document}

\title{Absolute Frequency Measurement of the 378 nm Transition in Thallium}
\date{\today}
\author{Tzu-Ling \surname{Chen}}
\affiliation{Department of Physics, National Tsing Hua University, Hsinchu, Taiwan 30013}

\author{Isaac \surname{Fan}}
\affiliation{Department of Physics, National Tsing Hua University, Hsinchu, Taiwan 30013}

\author{Chang-Yi \surname{Lin}}
\affiliation{Department of Physics, National Tsing Hua University, Hsinchu, Taiwan 30013}

\author{Hsuan-Chen \surname{Chen}}
\affiliation{Institute of Photonics Technologies, National Tsing Hua University, Hsinchu, Taiwan 30013}

\author{Shih-En \surname{Chen}}
\affiliation{Department of Physics, National Tsing Hua University, Hsinchu, Taiwan 30013}

\author{Jow-Tsong \surname{Shy}}
\affiliation{Department of Physics, National Tsing Hua University, Hsinchu, Taiwan 30013}
\affiliation{Institute of Photonics Technologies, National Tsing Hua University, Hsinchu, Taiwan 30013}

\author{Yi-Wei \surname{Liu}}
\email{ywliu@phys.nthu.edu.tw}
\affiliation{Department of Physics, National Tsing Hua University, Hsinchu, Taiwan 30013}

\begin{abstract}
The absolute frequency measurement for all of the hyperfine transitions of ${6P_{1/2}\rightarrow7S_{1/2}}$ using a self-referencing frequency comb is reported. This transition can be used as a bench marker for the accurate atomic wave function, and then improve the calculation of the atomic parity-non-conservation (PNC).
The frequency center is precisely determined by saturation spectroscopy utilizing a pair of the counter-propagating laser beams intersecting with atomic beam.
Such a Doppler-free profile has been improved to 350~kHz in the absolute frequency measurement. The $\rm{7S_{1/2}}$ state hyperfine splittings have also been deduced from our results.

\end{abstract}


\maketitle

\section{INTRODUCTION}
High-precision measurements in atomic system shows very promising in testing the standard model of elementary particles and searching for new physics beyond it. Thallium plays an important role in atomic parity non-conservation (PNC) and permanent electric dipole moment (EDM) experiment mainly attributing to its atom number, which enlarge PNC effect by $\rm{Z^3}$. In the most accurate atomic PNC experiment, an experimental accuracy of 0.35\% has been achieved in cesium~\cite{1997:Wood}, whose atomic structure is better understood. Combining the theoretical atomic-structure calculations only 0.5\% \cite{Dzuba:2002ta}, it leads to the weak charge $Q_{W}$ of cesium nucleus, which can be compared with the predication of the standard model. This result leads to the most accurate low-energy test of the electroweak sector of the SM to date. 
The PNC effect had also been observed in atomic thallium system using ${6P_{1/2}\rightarrow6P_{3/2}}$ transition in 1995 and reached 1.7\% of experimental uncertainty \cite{1995:Vetter, 1995:Edwards}, but the theoretical uncertainty for thallium is only at the level of 2.5\% \cite{2001:Kozlov,Dzuba:2006tk,Ginges:2004wa} due to the complicate atomic structure. The total uncertainty in the value of the inferred $Q_{W}$ is usually determined by summing experimental and theoretical uncertainties in quadrature. Thus, the resultant $Q_{W}$ of thallium nucleus is only with an accuracy of 2.8\% and limit the interpretation of the single-isotope PNC measurements.

In atomic thallium, there exists a strong correlation among the three valance electrons and hence it can not be treated accurately using the conventional many-body-perturbation theory (MBPT), as pointed out by Kozlov \cite{2001:Kozlov} and Dzuba \cite{1996:Dzuba}. 
A new theoretical approach that combines the MBPT with the configuration interaction (CI) and the coupled-cluster method has been developed to incorporate the correlation effect \cite{Safronova:2011cn}. 
To verify the accuracy of this new theoretical approach, however, various observables such as transition energies, hyperfine splittings, transition amplitudes, and polarizabilites, need to be compared to the corresponding experimental values \cite{safronova05-pra}. 

Among the lower lying energy levels of Tl, the $6P_{1/2}\leftrightarrow7S_{1/2}$ transition is of particular interest.
An accurate energy determination of this ground state transition not only can aid the development of the many-body theory, but it can also complement experiments where this transition is utilized, such as in the laser cooling of Tl \cite{Fan:2011bf}.
Hyperfine splittings and isotope shifts of this transition have been previously measured in the vapor gas cell \cite{2000:Majumder} and atomic beam \cite{1993:Hermann} setups.
The precision in both setups, however, are limited by the Doppler-broadened signal profiles.
The Doppler-broadening effect, in principle, can be eliminated by using the saturation spectroscopy technique \cite{2003:Demtroder, klimack84-pra, Tissue94-josab}.
In such case, the decrease (a dip) in the absorption strength at resonance is induced by the saturation effect and the dip profile can be interpreted as Doppler-free having a linewidth smaller or equal to the natural linewidth of the corresponding transition.  

The Doppler broadened spectrum of the ground state resonance of ${6P_{1/2}\rightarrow7S_{1/2}}$ transition of atomic thallium, the hyperfine splittings and transition isotope shift have been measured using both gas cell \cite{2000:Majumder} and atomic beam \cite{1993:Hermann} with the direct absorption and fluorescence technique. 

For Tl, the atomic beam setup should have a distinct advantage over the vapor cell setup in terms of observing the Doppler-free feature.
The three $6P_{1/2}\leftrightarrow7S_{1/2}\leftrightarrow6P_{3/2}$ levels of Tl constitute a $\Lambda$-type energy system where the saturation intensity of the $6P_{1/2}\leftrightarrow7S_{1/2}$ transition is significantly reduced due to the presence of the $6P_{3/2}$ metastable state \cite{Daily:1978du}, which traps the population with al long lifetime. In such a three-level ${\rm Lambda}$ system, the population
 is easily saturated and an additional homogeneous broadening (e.g. power broadening) can present in the signal profile.
This is especially true in the vapor cell setup where the same Tl atom is likely to absorb multiple photons before it travels outside of the interaction region.
In the atomic beam setup, on the other hand, a continuous influx of fresh Tl atoms prevents the lowering of the saturation intensity by limiting the time atoms reside in the interaction region.
The trade-off is the transit-time broadening, which is estimated to be less than 0.9 MHz if a 1 mm sized laser beam is shine onto the thermally activated atomic beam (243 m/s).

In this report, the saturation spectroscopy technique is applied to a Tl atomic beam to carry out a frequency comb measurement on all $6P_{1/2}\leftrightarrow7S_{1/2}$ hyperfine transitions without the Doppler-broadening effect.
The well-known saturation dip (``Lamb dip'') is observed and its absolute frequency is shown to be independent of the geometric relation between the optical beams (pump and probe) and the atomic beams under the condition that a \textit{collinearity} for the counter-propagating optical beams is maintained.
This removes the Doppler-shift of the frequency marker and hence alleviates a common systematic error that hinders the absolute frequency measurement in a atomic beam setup \cite{gerginov04-pra, hannemann06-pra, das07-pra, friebe08-pra, Salumbides:2011ts}.  

An optical frequency comb (OFC), which is a versatile optical frequency measuring tool with an accuracy of kHz, or better, was used in our experiment. Therefore, the precision of optical transition frequency measurement is only limited by the linewidth of the observed atomic or molecular transitions and signal-to-noise ratio of the spectrum, rather than the frequency measuring tool, as previous experiments. In addition, this is the first time the absolute frequency of the $6P_{1/2}\leftrightarrow7S_{1/2}$ transition in Tl is measured.
Along with the derived hyperfine splittings and isotope shifts, these measurements should provide invaluable information for calibrating the new many-body theory.

\section{EXPERIMENT}
\begin{figure}[t]
\includegraphics[width=1\linewidth]{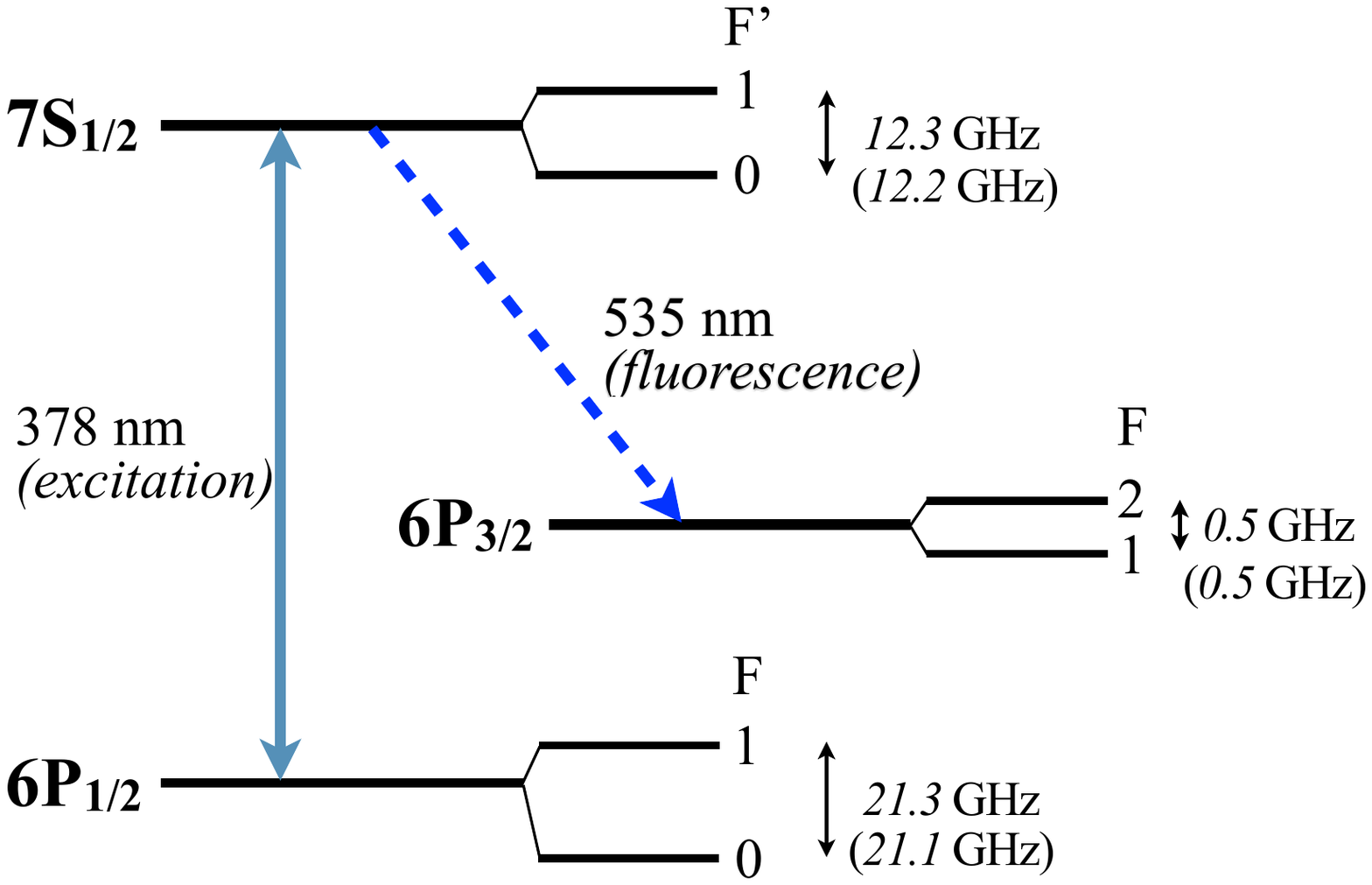}
\caption{\label{fig:energylevel}(Color Online) Simplified energy level diagram of $^{205}$Tl ( Tl$^{203}$).}
\end{figure}

The experiment setup is a typical pump-probe configuration with a thermal atomic beam. A schematic representation is shown in Fig.~\ref{fig:exp setup}. The 378~nm laser source was generated using a frequency doubled Ti:Sapphire laser, and splitted into pump and probe beams in counter-propagating directions. Both of them were linearly polarized and perpendicularly interacted with the atomic beam. To improve pumping efficiency and to take advantage of the long lifetime of the metastable 6{\rm $P_{3/2}$} state, the technique of optical pumping, as \cite{Klimcak:1984wt}, was employed. That is setting the pump beam to be in front of the probe beam, rather than overlapping each other. The population of ${\rm 6P_{1/2}}$ state was depleted, before probing.  The probe beam intensity was modulated using an optical chopper with a frequency of 2~kHz to reduce noise. The laser induce 535 nm fluorescence (${\rm7S_{1/2}\rightarrow6P_{3/2}}$) was detected using a photomultiplier, and demodulated using a lock-in amplifier.

\subsection{Apparatus}
\begin{figure}[t]
	\includegraphics[width=1\linewidth]{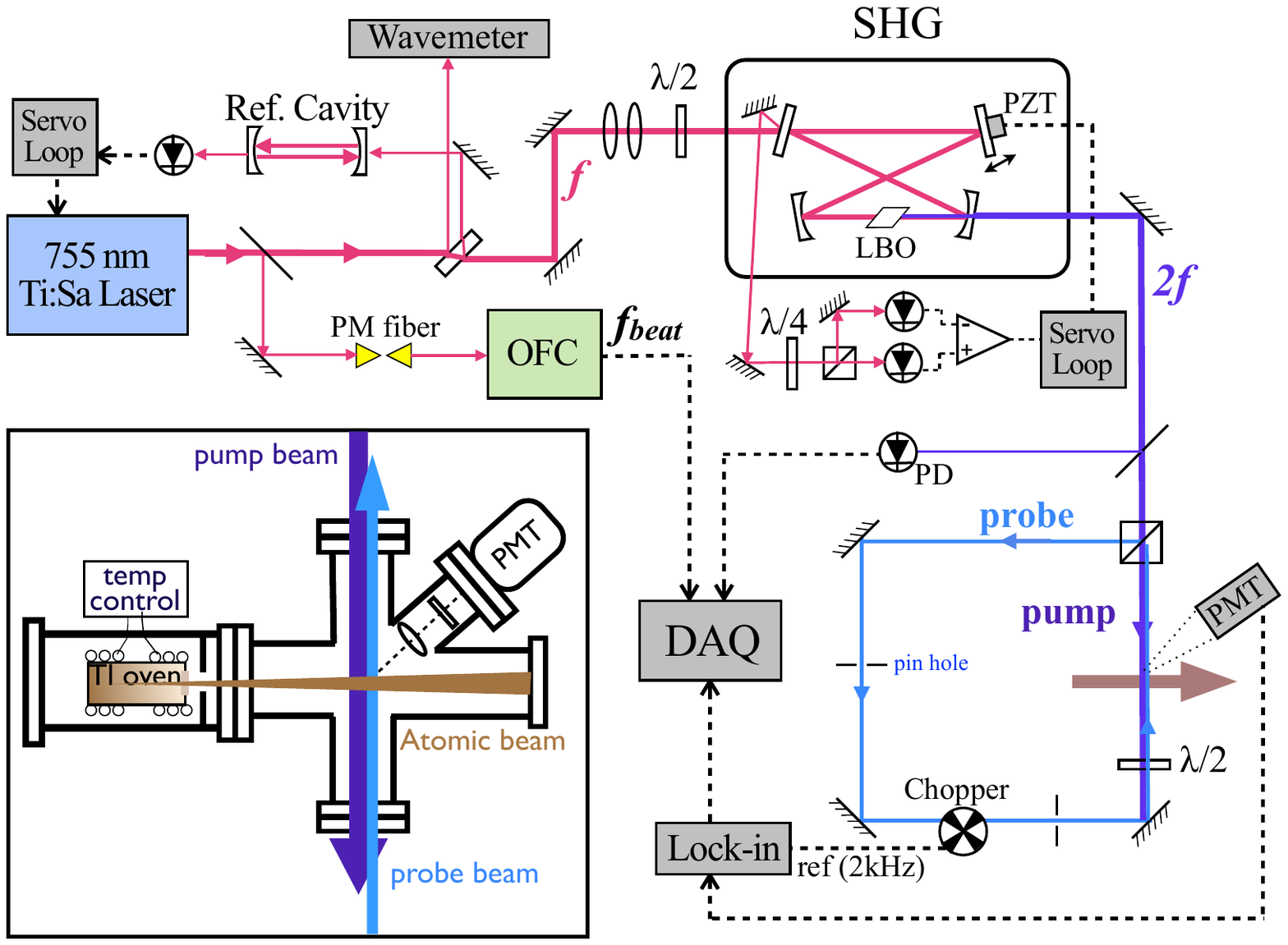}
	\caption{\label{fig:exp setup}(Color Online) The experimental setup of the absolute frequency measurement of thallium ${6P_{1/2}\rightarrow7S_{1/2}}$ transition. Inset:  The atomic beam apparatus.}.
\end{figure} 

A power of 1-2 mW of 378~nm UV light was generated using a LBO crystal in an enhancement cavity with an X-type ring configuration. The Cavity was locked to the fundamental laser frequency using the technique of polarization rotation. The detailed description of the system can be found in \cite{Fan:2011bf}. The laser beam was then collimated to a beam size of 3~mm${\rm \times}$ 10~mm using an AR coated lens. The elliptic beam shape was for a longer interaction time along the atomic beam propagating direction, therefore a higher pumping efficiency and a smaller transit-time broadening. 

The thermal atomic beam, shown in the inset of Fig.~\ref{fig:exp setup}, was generated from thallium bulk heated to $\rm{450^{\circ}C}$. It was then sieved out using 2~mm and 4.35~mm apertures separated by 12.85 mm, corresponding to a divergence of 90~mrad. The most probable velocity of thallium atoms was $\sim$243~m/s. The experimental chamber was pumped using a small turbo pump (50 l/s) to maintain a pressures below $10^{-5}$~Torr. The density of atoms in the beam is $\sim 10^{12}$ cm$^{-3}$.

The 755~nm fundamental laser pumped by a 10W 532~nm laser provided a power of $\sim$850~mW, and was locked to a reference cavity of invar alloy with a finesse of 310 and a free spectral range of 1.07~GHz. This cavity is also equipped with a piezo stack for frequency scanning. The frequency jitter was then reduced to ${\rm<1~MHz}$ in 0.1~sec, and the frequency drift was ${\rm<15~kHz/sec}$ due to the cabinet temperature change. A part of the 755~nm laser output was mixed with the OFC through a polarization maintained optical fiber for absolute frequency measurement.  The detailed information about this OFC system can be found in \cite{Lien:2011cb}.

\subsection{Spectrum and frequency measurement}
The 755~nm laser was delivered to the OFC system using a polarization-maintained fiber, which was used to avoid the fluctuation of the laser power due to the birefringence effect of an optical fiber. The 755~nm laser and OFC were then combined using a polarizing beam splitter (PBS), and the beat frequency ($f_{beat}$) which was detected using an avalanche photodiode (APD) and measured using a counter and a spectrum analyzer. For a reliable counting of the beat frequency, the signal of beat note was kept to be $>$30~dB above noise level.
 
The scan rate of 755 nm laser was 30~MHz/min. A single scan consists of 1000--1500 data points within the frequency range of 300 MHz in terms of the 755~nm laser frequency. Each data point is a set of information including: the beat frequency $f_{beat}$, OFC repetition rate $f_{beat}$, OFC offset frequency $f_{o}$, the averaged fluorescence signal, the standard deviation of the fluorescence signal and 378~nm laser power. The data collecting rate is 2~sets/sec. The time constant of the lock-in amplifier was 30~ms that is sufficiently short to follow such a slow scan rate.  Furthermore, during data taking, two "forward-backward" scans (4 scans) were performed to cancel out any possible "direction-dependent" systematic effect.

\subsection{Lineshape Model}
A typical line shape of the saturation dip with a Doppler-broadened background is a Lorentzian dip on the top of a Gaussian profile. However, it would be insufficient to fully describe our experimental observations, which are the mixing of the homogenous and the inhomogenous broadening effects and the dispersive asymmetry. To incorporate these features, in our experiment, the lineshape model is:
\begin{equation}
\label{eq:shape}
\begin{split}
\rm{S(\omega)} &=(\alpha+\rm A(\rm{\omega, \omega_{0}+\omega_{shift}}))\times\rm{V(\omega,\omega_0+\omega_{shift},\Gamma_{L1},\Gamma_{G})}\\
&-\beta\times\rm{L(\omega,\omega_0,\Gamma_{L2})}+c,
\end{split}
\end{equation}
The Voigt function, $\rm{V(\omega,\omega_0+\omega_{shift},\Gamma_{L1},\Gamma_{G})}$, is the convolution of the Gaussian and Lorentzian functions to describe the mixing of the homogenous-inhomogenous broadening. The Lorentzian function, $\rm{L(\omega,\omega_0,\Gamma_{L2})}$, is the homogeneously broadened, Doppler-free dip. The minus sign "-" of the second term indicates the "dip". ${\rm\Gamma_{L1}}$ and ${\rm\Gamma_{L2}}$ are the linewidths of the Lorentizan (homogenous) profiles, and the ${\rm\Gamma_{G}}$ is the Gaussian width (Doppler broadening, inhomogenous). $\omega_{0}$ is the center frequency of the transition. $\omega_{shift}$ is a small shift between the Voigt background and the Lorentzian dip, due to imperfect perpendicularity between the atomic beam and the lasers (see section~\ref{sec:uncert}).

$\rm A(\rm{\omega, \omega_{0}+\omega_{shift})}$ is a derivative Gaussian profile for the asymmetry of the background profile. The asymmetric feature has been observed in many spectroscopy experiments using the fluorescence or absorption techniques, and found to be induced by various mechanisms, including the atomic recoil effect in a dispersive medium~\cite{Gerginov200317, Gerginov:2006kx}, collision time effect \cite{HARRIS:1984uh} as well as the atomic recoil effect due to the velocity-dependent force~\cite{Prentiss:1986vl}. The asymmetry of the lineshape can be quantified using a single parameter $\epsilon$, the ratio of the maximum slopes of the two sides of the peak, as defined by \cite{Prentiss:1986vl}. $\epsilon$ is $\sim$0.85 in our work. For a small asymmetry, a derivative Gaussian profile is an good approximation \cite{FloresLlamas:2001kz}:

\begin{equation}
\label{eq:asymmetry}
\begin{split}
&\rm{A}(\omega, \omega_0+\omega_{shift})=\frac{d}{d\omega}\rm{Gaussian}(\omega,\omega_0+\omega_{shift},\Gamma)\\
&=a\times(\omega-(\omega_0+\omega_{shift}))\times\rm{Gaussian(\omega,\omega_0+\omega_{shift},\Gamma)}\\
\end{split}
\end{equation}
This model works well with our experimental results, as the fitting residuals show mainly random noise without any special pattern (see Fig.~\ref{fig:spectrum}). The ratio of asymmetry of our spectrum fitted to this model is ${\rm (a/\alpha)<2\%}$. The fitting program is written on ROOT platform (CERN) using the built-in Voigt function.

\section{RESULTS AND DISCUESSION}

\subsection{Doppler-free spectrum}

\begin{figure}[t]
	\includegraphics[width=1.0\linewidth]{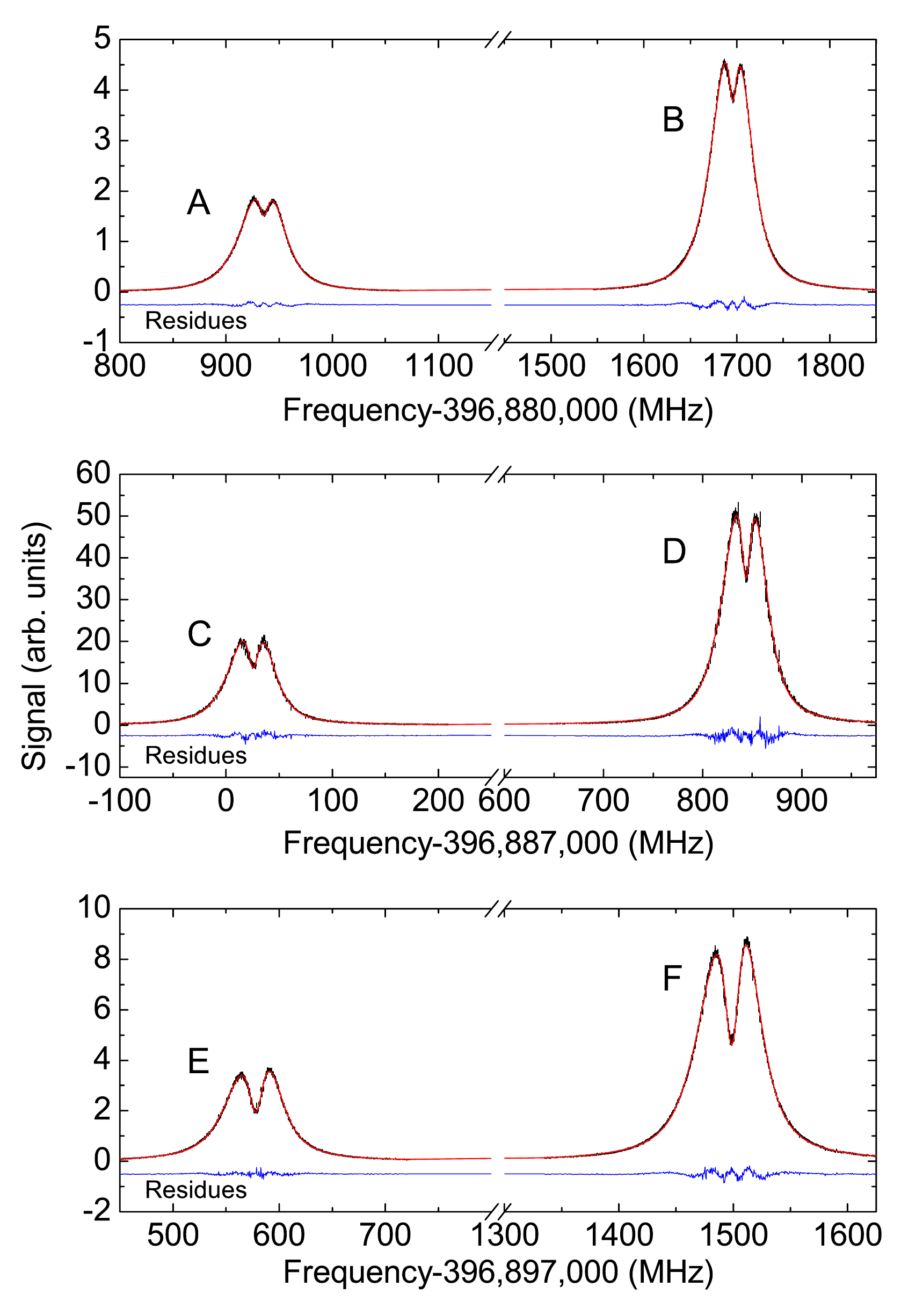}
	\caption{\label{fig:spectrum}(Color Online) The spectrum of all hyperfine transitions. The fitting line (red solid curve) and fitting residues (blue curve) are also shown. The frequency axis shown in this figure is the frequency of the fundamental 755~nm laser. ${\rm A:[^{203}Tl]~6P_{1/2}(F=1)\rightarrow7S_{1/2}(F=0)}$, ${\rm B:[^{205}Tl]~6P_{1/2}(F=1)\rightarrow7S_{1/2}(F=0)}$, ${\rm C:[^{203}Tl]~6P_{1/2}(F=1)\rightarrow7S_{1/2}(F=1)}$, ${\rm D:[^{205}Tl]~6P_{1/2}(F=1)\rightarrow7S_{1/2}(F=1)}$, ${\rm E:[^{203}Tl]~6P_{1/2}(F=0)\rightarrow7S_{1/2}(F=1)}$, ${\rm F:[^{203}Tl]~6P_{1/2}(F=0)\rightarrow7S_{1/2}(F=1)}$}
\end{figure}

Fig.\ref{fig:spectrum} shows all the hyperfine transitions of ${6P_{1/2}\rightarrow7S_{1/2}}$ in Tl with fitting curves and fitting residues.  Each spectrum is histogram composed of four individual scans with 500~kHz bin size, which was chosen to be efficient and without losing any accuracy. The errorbars of histogram were given by the deviation of the fluorescence signal strength within the same laser frequency bin. The signal-to-noise ratio (SNR) is 35-50, which allows a sub-MHz accuracy in measuring the center frequency of a dip with a 13~to~20~MHz width.

The dip widths of ${6P_{1/2}(F=1)\rightarrow 7S_{1/2}(F'=0)}$ (A and B) are only $\sim13$ MHz and smaller than the other hyperfine transitions to the F'=1 excited states, which are C, D, E and F lines with dip widths ${\rm\sim20~MHz}$. This is due a larger branching decay rate of ${\rm 7S_{1/2}(F=1)}$ to the metastable ${\rm 6P_{3/2}}$ state, and causes a lower effective saturation intensity and a stronger power broadening. The widths of background Voigt profiles are 30-50~MHz contributed from the nature linewidth, the first order Doppler broadening, and the power broadening. All of the laser linewidth ($<$1~MHz), the transit-time broadening ($<$0.1~MHz), the collision broadening ($<$0.1~MHz) are negligible in our experiment.

\begin{figure}[t]
	\includegraphics[width=1.0\linewidth]{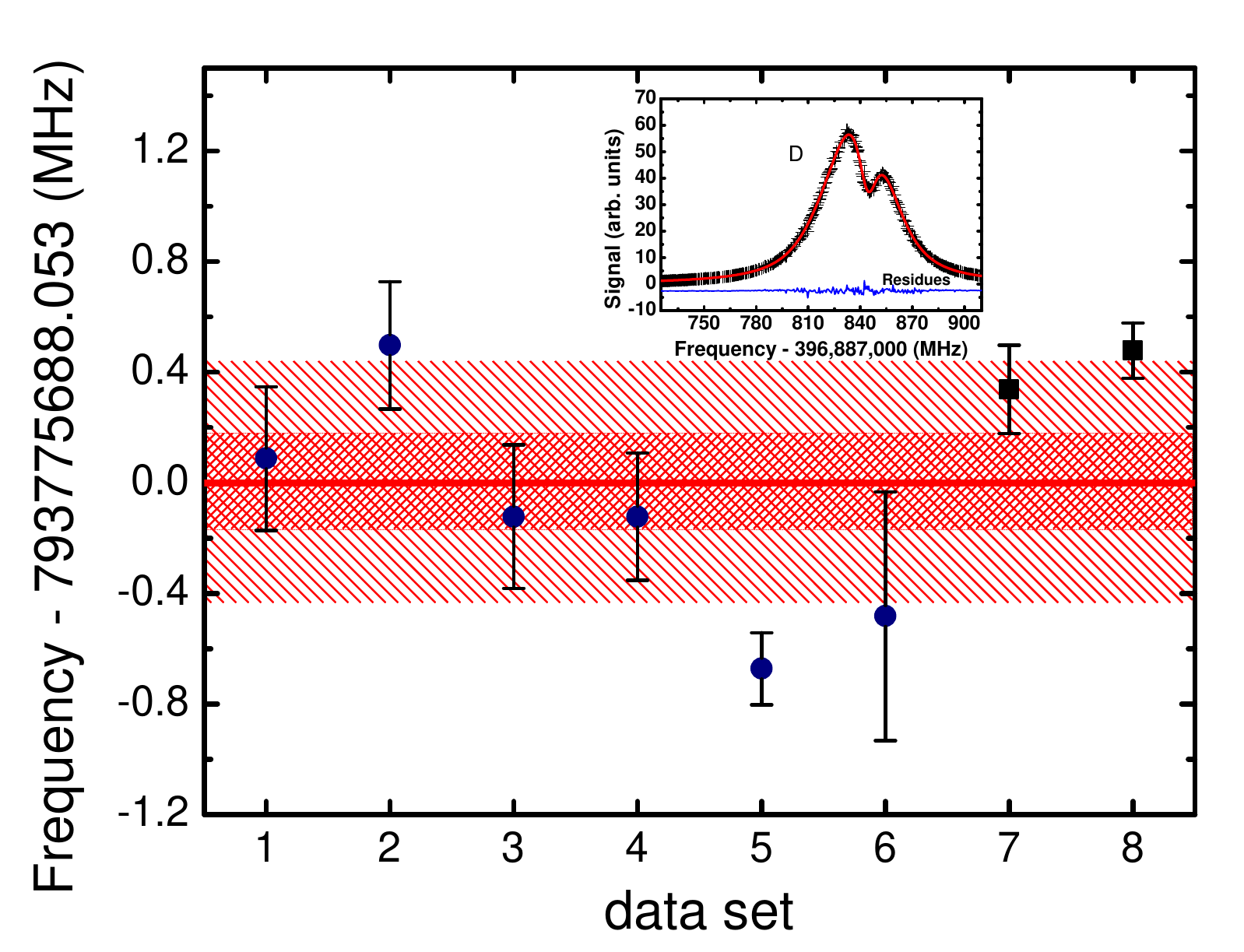}
	\caption{\label{fig:misalign}(Color Online)The scattering distribution of the D-line absolute frequency measurements. Two shadows sections indicate the regions of the standard error of the mean (${\rm\pm0.18~MHz}$) and the ${\rm 1\sigma}$ standard deviation (${\rm\pm0.45~MHz}$) of the distribution. The $\bullet$ points are the measurements with good perpendicularity, and the $\blacksquare$ points are with large deviations from the perpendicular position (see text). The inset shows a typical line shape of the $\blacksquare$ points.}
\end{figure}

\subsection{Frequency uncertainty assessment}\label{sec:uncert}
The uncertainties contributed from various sources, including the statistical and the systematic, are summarized in Table~\ref{table:error}.

The statistic uncertainty of the center frequency is 0.12 to 0.22~MHz by fitting a single spectrum composed of four scans to the line shape model. It was limited by the linewidth and the SNR, which is attributed to the laser intensity noise (${\rm<5\%}$), the fluctuations of the atomic beam, the laser beam pointing vibration, the electronic noise, and so forth. For the long-term drift, a day-to-day stability has been tested using the D line. As shown in Fig.~\ref{fig:misalign}, the standard error of the mean was found to be 0.18~MHz, which is similar to the uncertainty given by a single spectrum statistic. For all the transition lines, the final statistic uncertainties were given to be 0.22~MHz.

The systematic errors is  mostly contributed by the residual first-order Doppler shift. Ideally, the saturation spectroscopy with collinear lights interacting with an atomic beam would strongly suppress such a shift of the saturated dip, which is resulted from the stationary atoms. However, in an atomic beam experiment, in order to measure a frequency to a high precision, two systematic effects must be carefully examined: the collinearity between the pump and the probe beams, and the perpendicularity between the lasers and the atomic beam. 

The collinearity was estimated to be better than 0.43~mrad, i.e., a 0.5~mm separation of the laser beams in a 600~mm distance, which causes a shift of 270~kHz. This uncertainty was limited by the visibility of two divergent UV laser beams in a far distance.

In a single laser beam excitation experiment,  the non-perpendicular intersection of laser and atomic beam results a shift of center frequency, due to the first order Doppler effect.  In our experiment, this problem was resolved by using a pair of counter-propagating beams-pump and probe. It is known that the non-perpendicularity will not affect the position of the dip, but only the fluorescence background. However, the relative shift between the dip and the background (${\rm \omega_{shift}\neq0}$), as the inset of Fig.~\ref{fig:misalign}, will induce a strong asymmetry in the spectrum profile, and could still possibly shift the center frequency. In order to examine such a effect, both the two counter-propagating laser beams were purposely adjusted to be largely deviated from the perpendicular intersection.  A large angle between lasers and atomic beam is set to be $\rm{ \Delta\phi\sim20~mrad}$, and enlarges the ${\rm\omega_{shift}}$ to ${\rm \sim 6~MHz}$ (the inset of Fig.~\ref{fig:misalign}). However, the center frequencies ${\rm\omega_0}$ of these misaligned measurements ($\blacksquare$ in Fig.~\ref{fig:misalign}) were found to be still within ${\rm1~\sigma}$ deviation of the measurements with careful alignment to the perpendicularity. These measurements demonstrates the Dopple-free feature of the saturated dip.

The relativistic second-order Doppler shift, for the thermal atomic beams with velocities of 243 m/s, is only 0.3 kHz.
The light shifts (ac Stark shifts) is negligible in our experiment, because there is no energy level, which is linked to the relevant states (${6P_{1/2}}$ and ${7S_{1/2}}$), introducing substantial perturbation by the 378~nm excitation laser. As for the Zeeman effect, while the laser was linearly polarized in the direction perpendicular to the 0.5~G Earth filed and the atomic beam, it mainly broadens the linewidth, rather than shifts the line center. However, due to the finite extinction ratio of the polarizer, the shift was estimated to be ${\rm<5~kHz}$. The uncertainty due to OFC is $\rm{\sim}$20 kHz, as discussed in \cite{Lien:2011cb}.

\begin{table}
\caption{\label{table:error}Budget of corrections and uncertainties of absolute transition frequency in Tl $6P_{1/2} \rightarrow 7S_{1/2}$ measured on a thermal atomic beam: statistical results and systematic error budget. All the values are in MHz.}
\begin{center}
\begin{tabular}{c|l}
Source & uncertainties \\
\hline\hline
Statistics & 0.22\\
frequency comb& 0.04\\
Residue Doppler shift  & 0.27\\
Second-order Doppler  & 0.0003\\
Zeeman effect  & 0.005\\
\hline
Total & 0.35\\
\hline\hline
\end{tabular}
\\
\end{center}
\end{table}

\subsection{The absolute transition frequencies}
The absolute transition frequency was obtained by measuring the fundamental infrared 755~nm laser, then times two. For each data point, three frequencies (${\rm f_{rep}}$, ${\rm f_{o}}$, ${\rm f_{beat}}$) were directly measured using counters, and recorded. The absolute frequency is then calculated using the following simple equation: 
$$f_{measure}=N\times f_{rep}\pm f_{o}\pm f_{beat}\ ,$$ 
where the repetition rate of comb laser $f_{rep}$  is ${\rm \sim1~GHz}$, the offset frequency of comb laser $f_{o}$ and the beat frequency $f_{b}$ were typically several hundreds MHz. The ${\rm \pm}$ signs and the $N$ (a large integer, $\sim 4\times10^{5}$ for 755~nm) were firstly inferred from the wavelength measurements of a 0.5~GHz accuracy wavemeter  and then confirmed using the method described in \cite{Lien:2011cb}. 

The final absolute frequencies of all six components of thallium ${6P_{1/2}\rightarrow7S_{1/2}}$ transition, including three hyperfine transitions and two isotopes, are listed in Table~\ref{table:freq}.  They are labeled as $\rm{A-F}$ for identification, and in comparison with the previous experimental measurements and the theoretical calculations. The center-of-gravity value of $^{205}$Tl calculated from the absolute frequency is also listed to be compared with the previous determination of \cite{Kurucz, 2001:Kozlov}.

\begin{table}
\caption{\label{table:freq}The transition frequencies of ${6P_{1/2}(F)\rightarrow7S_{1/2}(F')}$.}
\begin{center}
\begin{tabular}{c|cl}
Line & $F-F'$ & Laser frequency (MHz)\\
\hline\hline
A & $\rm{^{203}Tl}$ $\rm{,1-0}$ & 793~761~871.36(35)\\
B  & $\rm{^{205}Tl}$ $\rm{,1-0}$ & 793~763~391.29(35)\\
C & $\rm{^{203}Tl}$ $\rm{,1-1}$ & 793~774~051.26(35)\\
D & $\rm{^{205}Tl}$ $\rm{,1-1}$ & 793~775~687.38(35)\\
E  & $\rm{^{203}Tl}$ $\rm{,0-1}$ & 793~795~156.48(35)\\
F  & $\rm{^{205}Tl}$ $\rm{,0-1}$ & 793~796~997.62(35)\\
\hline\hline
cg (of Tl$^{205}$)\\
\hline
this work &${6P_{1/2}\rightarrow7S_{1/2}}$&793~777.941(1)~(GHz)\\
Experiment \cite{Kurucz}&${6P_{1/2}\rightarrow7S_{1/2}}$&793~775.5~~~~~~~~(GHz)\\
Theory\cite{2001:Kozlov}&${6P_{1/2}\rightarrow7S_{1/2}}$&793~100(10)~~~~~(GHz)\\
\end{tabular}
\end{center}
\end{table}

\subsection{HFS, IS and the Mean Square Isotopic Change $\lambda_{c,m}$}

\begin{table*}
\caption{Summary of measurements of thallium ${6P_{1/2}}$ and ${7S_{1/2}}$ hyperfine constants A and isotope shift (IS) for two stable isotopes. All results are in MHz. The last three columns present theory comparison.}
\begin{ruledtabular}
\begin{tabular}{clllll}
\label{table:HFS}
&${6P_{1/2}}$~$^{205}$Tl  & ${6P_{1/2}}$~$^{203}$Tl &${7S_{1/2}}$~$^{205}$Tl & ${7S_{1/2}}$~$^{203}$Tl & IS of ${6P_{1/2}\rightarrow7S_{1/2}}$\\
\hline\hline
This work 	& 21310.24(69) 	& 21105.22(69) 	& 12296.09(69) 	& 12179.90(69) 	& 1658.33(69)\\
Ref.\cite{2000:Majumder}		&  			& 			& 12294.5(15) 	& 12180.5(18) 	& 1659.0(6)\\
Ref.\cite{1993:Hermann}		& 			& 			& 12297.2(16)  	&12181.6(22)  	&   \\
Ref.\cite{1985:Neugart}		&   			&    			& 12284.0(60)  	&  12172.0(60) 	&    \\	
Ref.\cite{1962:Shuler}			&      			&           		&  12318(36) 	&  12225(42)    	&  \\
Ref.\cite{1956:Lurio}			& 21310.835(5)& 21105.447(5) &  			& 			& \\
\hline
Theory\cite{2001:Kozlov}&21663&&12666&&\\
Theory\cite{1995:Mrtensson-Pendrill}&21300&&12760&\\
Theory\cite{1998:Dzuba}&21623&&12307&\\
% Lines of table here ending with \\
\end{tabular}
\end{ruledtabular}
\end{table*}

The hyperfine splittings of the $6P_{1/2}$ ground state and the $7S_{1/2}$ excited state can be deduced from our measurements as listed in Table~\ref{table:HFS}. For the  ground state, our results are in good agreement with previous measurements, i.e., the radio-frequency measurement\cite{1956:Lurio}, with less accuracy. For the $7S_{1/2}$ excited sate, the hyperfine splitting has been improved by a factor of 2, in comparison with the latest measurement \cite{2000:Majumder}. 

There are two naturally occurring isotopes of thallium, $\rm{^{205}Tl~(70.5\%)~and~^{203}Tl~(29.5\%)}$. The isotope shift (IS) of the ${\rm 6P_{1/2}\rightarrow7S_{1/2}}$ can only be measured using optical spectroscopy. Our results are in good agreement with \cite{2000:Majumder}, which employed a gas cell and an optical reference cavity, in comparable accuracies.  

\begin{figure}[t]
	\includegraphics[width=1.0\linewidth]{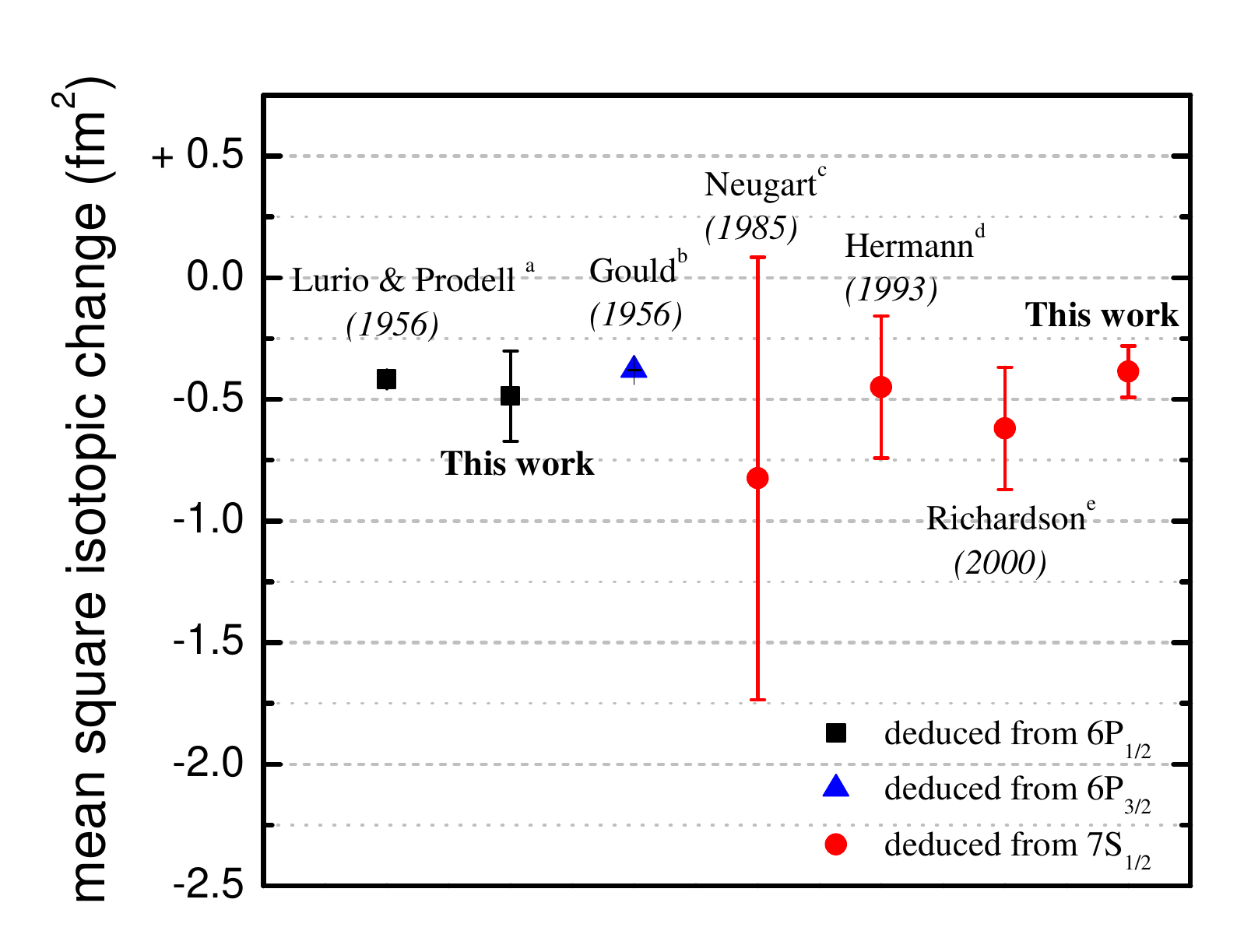}
	\caption{\label{fig:change}(Color Online) Comparison of the mean square isotopic change $\lambda_{c,m}$ based on various experimental results.\\
	$\lambda_{c,m}$ values shown in the figure were extracted by combining theoretical and experimental values. The value $\rm{\Delta/\lambda_{c,m}=-2.48\times10^{-4}~fm^2}$ for ${6P_{1/2}}$,  43.0${\times10^{-4}}$~$\rm{fm^2}$ for $\rm{6P_{3/2}}$ and ${-7.62\times10^{-4}~\rm{fm^2}}$ for ${7S_{1/2}}$ were theoretically calculated by \cite{1995:Mrtensson-Pendrill}. We infer $\lambda_{c,m}=-0.386(94)\ \rm{fm^{2}}$ in this work. This parameter inferred from the results of the ${6P_{1/2}}$, ${6P_{3/2}}$ and ${7S_{1/2}}$ state HFS are shown here.\\
$^{a}$ Ref.\cite{1956:Lurio}, $\lambda_{c,m}=-0.418(1)\ \rm{fm^{2}}$ of $\rm{6P_{1/2}}$.\\
$^{b}$ Ref.\cite{Gould:1956uj}, $\lambda_{c,m}=-0.379(2)\ \rm{fm^{2}}$ of $\rm{6P_{3/2}}$.\\
$^{c}$ Ref.\cite{1985:Neugart}, $\lambda_{c,m}=-0.825(910)\ \rm{fm^{2}}$ of $\rm{7S_{1/2}}$.\\
$^{d}$ Ref.\cite{1993:Hermann}, $\lambda_{c,m}=-0.450(292)\ \rm{fm^{2}}$ of $\rm{7S_{1/2}}$.\\
$^{e}$ Ref.\cite{2000:Majumder}, $\lambda_{c,m}=-0.620(251)\ \rm{fm^{2}}$ of $\rm{7S_{1/2}}$.\\
}
\end{figure} 

The mean square isotopic change ${\rm\lambda_{c,m}}$ between ${\rm^{203}Tl}$ and ${\rm^{203}Tl}$,  which is sensitive to both the radial distribution of the neutron magnetization and charge distributions in the isotopes, can be deduced from our results by calculating the hyperfine anomalies of the ${\rm6P_{1/2}}$ and ${\rm7S_{1/2}}$ states. Both of two measurements should give the same value of the isotopic change.
The hyperfine anomaly is written as , $\Delta=[(A{^{205}}/A{^{203}})(\mu_{I}^{203}/\mu_{I}^{205})-1]$, where $A{^{205}}$ and $A{^{203}}$ are the hyperfine constants and can be found in Tab.\ref{table:HFS}. Using the known ratio between the nuclear magnetic moments $\rm\mu_{I}(^{205}Tl)/\mu_{I}(^{203}Tl)$~=~1.009~836~13(6) measured in~\cite{Baker:1963fj}, ${\rm\Delta_{7S_{1/2}}=-2.9(7)\times10^{-4}}$ and ${\rm\Delta_{6P_{1/2}}=-1.21(46)\times10^{-4}}$ were resulted. To infer the mean square isotopic change, the ratio of $\rm{\Delta/\lambda_{c,m}}$ of the atomic thallium has been theoretically calculated by \cite{1995:Mrtensson-Pendrill}, which gave $\rm{\Delta/\lambda_{c,m}=-7.62\times10^{-4}~fm^2}$ for ${7S_{1/2}}$, and $\rm{-2.48\times10^{-4}~fm^2}$ for ${6P_{1/2}}$. Therefore, 
 $$\lambda_{c,m}=-0.39(11)\ \rm{fm^{2}}{\rm(7S_{1/2})}$$ 
 $$\lambda_{c,m}=-0.49(19)\ \rm{fm^{2}}{\rm(6P_{1/2})}$$

Fig.~\ref{fig:change} shows the comparison of the mean square isotopic change $\lambda_{c,m}$ based on various experimental results. Among the experiments based on $\rm{7S_{1/2}}$ state, our result gives the most precise value and is in very good agreement with the results deduced from the most accurate experimental values based on $\rm{6P_{1/2}}$ and $\rm{6P_{3/2}}$ states. Meanwhile, the value haven by our  $\rm{6P_{1/2}}$ state measurement is consistent with our own $\rm{7S_{1/2}}$ state measurement and in agreement with the others, although the uncertainty is considerably larger than \cite{1956:Lurio}.

\section{Conclusion}
\label{sec:conclusion}
The ${6P_{1/2}\rightarrow 7S_{1/2}}$ transition frequencies of Tl were measured by Doppler-free saturation spectroscopy with an optical frequency comb (OFC) in an atom beam configuration, where the saturated dip had been verified to be shift-free even if the lights are not perpendicular to atomic beam. For the first time, the absolute transition frequencies of six components of $\rm{^{203}Tl}$ and $\rm{^{205}Tl}$ were measured to a sub-MHz accuracy.

Meanwhile, the HFS and IS of ${6P_{1/2}}$ and ${7S_{1/2}}$ are determined by the absolute transition frequency values with an accuracy of 0.7~MHz. In which the HFS within ${6P_{1/2}}$ is in good agreement with, but of less precision to, the most accurate measured values by microwave techniques by Ref.~\cite{1956:Lurio}. However, we have made a greatly improved measurement within the ${7S_{1/2}}$ transition from the precision of MHz to sub-MHz. The achieved accuracy of the derived HFS and IS can provide more input data for the calculation of the PNC amplitude of thallium to improve the calculation accuracy~\cite{2001:Kozlov}.

The mean square isotopic change $\lambda_{c,m}$ deduced from the hyperfine anomalies of ${7S_{1/2}}$ state had also been improved. Our result is in very good agreement with the most precise value from the hyperfine anomaly of ${6P_{1/2}}$. This result provide one of the few handles on the neutron radial distribution in nuclei and calibration for nuclear structure calculations which are of importance to the understanding of PNC and QED effects in thallium and further heavy atoms~\cite{1995:Mrtensson-Pendrill,Grossman:1999ub}.

\begin{acknowledgments}
This work was supported by the National Science Council of Taiwan under Grant No.95-2112-M-007-005 and No. 94-2112-M-007-011.
\end{acknowledgments}
\bibliographystyle{apsrev4-1}
\bibliography{tl_ref}

\end{document}